\documentclass[
]{ceurart}

\usepackage{listings}
\begin{document}

\copyrightyear{2024}

\copyrightclause{Copyright for this paper by its authors.
  Use permitted under Creative Commons License Attribution 4.0
  International (CC BY 4.0).}

\conference{PhysioCHI: Towards Best Practices for Integrating Physiological Signals in HCI,   May 11, 2024, Honolulu, HI, USA}

\title{Implicit gaze research for XR systems}

\author{Naveen Sendhilnathan}[%
email=naveensn@meta.com,
]
\address{Reality Labs Research, Meta Platforms, WA, USA}

\author{Ajoy S. Fernandes}[%
email= ajoyferns@meta.com,
]

\author{Michael J. Proulx}[%
email=michaelproulx@meta.com,
]

\author{Tanya R. Jonker}[%
email=tanya.jonker@meta.com,
]

\begin{abstract}
  Although eye-tracking technology is being integrated into more VR and MR headsets, the true potential of eye tracking in enhancing user interactions within XR settings remains relatively untapped. Presently, one of the most prevalent gaze applications in XR is input control; for example, using gaze to control a cursor for pointing. However, our eyes evolved primarily for sensory input and understanding of the world around us, and yet few XR applications have leveraged natural gaze behavior to infer and support users' intent and cognitive states. Systems that can represent a user's context and interaction intent can better support the user by generating contextually relevant content, by making the user interface easier to use, by highlighting potential errors, and more. This mode of application is not fully taken advantage of in current commercially available XR systems and yet it is likely where we'll find paradigm-shifting use cases for eye tracking. In this paper, we elucidate the state-of-the-art applications for eye tracking and propose new research directions to harness its potential fully.
\end{abstract}


\maketitle

\section{Introduction}

The integration of eye-tracking technology into virtual reality and mixed reality (XR) headsets is becoming increasingly common. However, the extent of the role of eye tracking in these systems has yet to be discovered, and, likely, its greatest potential has yet to be realized. 

Currently, a popular use of gaze applications in XR systems revolves around using it as an explicit input control mechanism \cite{fernandes2023leveling}. For example, eye tracking might be used to control a cursor for object targeting, which may be visible or invisible to the user. Although eye-tracking-based input has been used in many interactions and interfaces, this approach presents various limitations which could result in fatigue, both physically from moving their eyes unnaturally, and mentally, as users exert willpower, going against how they would naturally attend to objects to maintain focus. Although improvements in explicit input control will continue to unlock opportunities in interactions, the number of applications might be limited relative to the opportunities presented when using eyes consistently with how they naturally move and how they have evolved.

Perhaps the most promising domain for the transformative value of eye tracking for XR lies in \textit{inferring user interaction intent and state}.  Users' natural gaze behavior provides a uniquely promising "implicit" signal to a user's internal state. Indeed, spatiotemporal gaze dynamics have been shown to correlate strongly with various cognitive and motor processes, including visual attention \cite{pastukhov2010rare}, motor preparation and execution \cite{de2018keeping, Naveen2021}, user confidence \cite{vine2016integrative, gupta2023}, decision-making \cite{gidlof2013using}, frustration \cite{stone2023unconscious}, and performance monitoring \cite{Naveen2022}. Therefore, while utilizing gaze as a controller might be a necessary starting point to catalyze explorations, a fruitful and vast research opportunity involves exploring how to add value to end users with help from their natural gaze to infer their intended future actions and cognitive states. This information can significantly enhance user experience, for example, by enabling personalization, content recommendations, adaptive guidance, and more. For example, the system might automatically bring up weather information when the user looks outside their window before they head out to work, or it might simply bring up the definition of a word if the user fixates steadily on that word within a blurb of text. However, the largely untapped potential of understanding and utilizing natural gaze behavior in current commercially available XR systems creates a gap in realizing the full capabilities of this technology.

In this workshop paper, we first provide a detailed definition of the two modes of gaze interactions in XR (explicit and implicit). We then provide a comprehensive discussion about the current state-of-the-art  applications, highlighting their advantages and limitations. Finally, we propose suggestions and recommendations for advancing eye-tracking research and outline the necessary future work to establish an XR interaction platform where eye tracking serves as a significant modality, enhancing user experience and facilitating effortless XR interactions.

\section{Modes of Gaze  Interaction in XR Systems} 

Gaze can be used in two different modes of interaction \cite{plopski2022eye}. First, gaze can be used in the  “explicit mode”, where one must control their eye movements to achieve a particular outcome. For example, a user might direct their eyes to particular interactable elements of a UI, so that they can navigate and manage the UI (e.g., using gaze position for cursor control or using dwell time and gaze gestures for selection \cite{fernandes2023leveling}). Alternatively, gaze can be used in the "implicit mode", where natural eye movements are preserved and can reflect a user's internal cognitive and emotional state. “Implicit” behaviors capture intentions, surprise, cognitive load, strain, learnability, and skill level.

The main differences between the explicit and implicit modes of interactions are summarized in Table 1 and will be expanded in detail in the following sections.

\begin{table*}[h!]
  \caption{Explicit and Implicit Modes of Interaction}
  \label{tab:interactions}
  \begin{tabular}{|p{1.3in}|p{2.3in}|p{2.3in}|}\hline
    \toprule
    &Explicit gaze interactions&Implicit gaze interactions\\
    \midrule
    
    Mode of UI control & Eye gaze & Other modalities such as hand\\\hline
    
    Primary use of gaze & Pointing, selecting & Visual search, passive gaze grazing etc \\\hline
    
    Distinct signatures of interaction & 
    Specific gaze features such as saccade vigor, velocity might be higher due to enhanced goal-directed movements. Only overt attention is used. & 
    Both overt and covert attention are utilized \\\hline

    Advantages & 
    Can be used to make high bandwidth interaction based selections quickly.
    Can be used when hands are occupied / unable to provide input.
     &
    User's natural gaze behavior is used to infer intentions, surprise, cognitive load, strain, learnability, skill level etc.  \\\hline
    
    Limitations & 
   
    Physical load of using eyes explicitly is high and might cause eye fatigue with prolonged use.
     Often suffers from Midas Touch problem and needs high-quality design to bypass it.
    Inability to use eyes to survey environment, spot dangers, etc. &
    Degeneracy of gaze behavior:  Same / similar implicit gaze dynamics could be picked up by different machine learning models to decode different states. 
    \\\hline

  \bottomrule
\end{tabular}
\end{table*}

\section{Applications of Explicit Mode of Gaze Interactions}

Gaze interactions allows users to interact with elements in their field of view by explicitly looking at them and selecting them. Some examples of this type of interaction include typing on keyboards by looking at the keys and pressing a button \cite{rajanna2018gaze} or using gaze as a cursor to navigate a UI and select elements using fixations or other gaze gestures \cite{khamis2018vrpursuits} and redirect stylus input to gaze location \cite{pfeuffer2015gaze}. The complexity of these interactions primarily depends on the density of UI elements, the coverage required for the user population and the eye tracker hardware capabilities such as accuracy, precision, depth estimate, frame rate, latency and also the types of interactions that the system is designed to achieve. 

Design choices, such as refining gaze targeting, determining the characteristics of interactable objects like item dimensions, padding, and spacing, as well as defining interaction techniques and providing visual feedback, can effectively tackle specific user behaviors and offset inaccuracies in eye tracking \cite{feit2017toward}. For instance, instead of strictly requiring the gaze cursor to be directly on a button for selection, one can utilize a margin between items. This can further be made dynamic based on context. For example, dynamically increase the margin size of certain objects that are most likely the target objects based on user history, time, etc. Additionally, providing visual feedback, such as highlighting buttons, has the potential to enhance the user experience and increase efficiency.

\section{Applications of Implicit Mode of Gaze Interactions}
Gathering real-time gaze information throughout different temporal stages, including before, during, and after an interaction, enables the inference of users' cognitive and behavioral states. Combined with the temporal axis, we explore two areas of research: 1) using implicit gaze for inferring interaction intent, and 2) leveraging implicit gaze for inferring the user's cognitive state.

\subsection{Using Implicit Gaze to Predict the Time and the Target for Point and Select Interaction}
Monitoring user gaze is a potential tool for evaluating design usability and enhancing workflow efficiency. In interactions involving multiple modalities, the combination of implicit gaze behavior and preceding hand movements, as demonstrated by David-John et al. \cite{david2021towards}, reliably predicts user's intended interactions. Such machine learning models that leverage gaze information to forecast user intent, assisting in tasks such as highlighting objects for selection and reducing false positive activations, thus mitigating physical strain and fatigue.  While such approaches, including the utilization of multimodal biosignals to predict target locations and interaction intent, are feasible, their accuracy depends on contextual features. For example, in the context of grocery shopping, knowledge of aisle locations informs potential targets, and item features aid in the selection process. A comprehensive model would take into consideration multimodal biosignals, context, and semantics, enabling the precise prediction of upcoming targets with high precision few tens of milliseconds or even seconds before user indication.

\subsection{Using Implicit Gaze to Enhance Interaction Primitives}
XR interactions like swipe typing, touch typing, scrolling, pointing, and selecting often demand significant physical and mental effort to be accurate. Leveraging contextual signals and implicit gaze can enhance user experiences, particularly in text entry and UI navigation. Using implicit gaze for wristband typing improves signal quality and reduces the need for precise input. In gesture typing for example swipe typing  \cite{zhao2023gaze}, aligning cursor speed with gaze fixation reduces hand movement by 30\% without compromising speed or accuracy. Dynamic acceleration adjustment near gaze fixation reduces hand movement by 22\%. Low effort wrist input, enabled by Flexible Pointing, Guidelines, and Magnets, facilitates quick interactions in a 2D UI during concurrent activities. Enhancements like gaze-based Magnets and scanpath history anticipate future targets, improving pointing precision and reducing interaction friction.

\subsection{Using Implicit Gaze to Detect Errors After User Interaction}
Despite the rapid advancements, the interaction with XR systems is still in its early stages of development due to ‘interaction errors’ that occur when there is a discrepancy between the system's response to a user's action and the user's anticipated outcome of that action. These errors usually arise from limitations in sensing, gesture recognition, and system decision-making. For instance, when a user attempts to pick up a pen using their index finger and thumb, the system may erroneously interpret it as a pinch gesture, resulting in unexpected system behaviors. These errors have a detrimental impact on user experiences as they cause frustration, necessitate manual corrections, and decrease trust in the system. In severe cases, they can even lead to users abandoning the system altogether. Recent research has shown that natural gaze can be used to detect user and system errors as early as 500-600 ms after their occurrence \cite{peacock2022gaze, Naveen2022} in a point-and-select interaction task.  Further research is necessary to build models that can generalize across different interaction methods and recognize different types of user and system errors. Such rapid detection of interaction errors can decrease the cognitive and motor loads of users and improve the usability of systems.

\subsection{Gaze Metrics To Estimate Users' Cognitive State}
In addition to using implicit gaze signals for user interactions, they can also serve to estimate a user's cognitive processes, which are not otherwise easily accessible. Gaze metrics derived from spatial and temporal gaze patterns are strong correlates of cognitive and motor processes such as visual attention \cite{frischen2007}, motor preparation and execution \cite{Naveen2021}, decision making \cite{reppert2015}, performance monitoring \cite{Naveen2022}, and more. Therefore, implicit gaze behavior is a valuable modality for investigating and measuring the cognitive state of the user while performing a task. 

Presently, manual adjustments are necessary in MR to determine the visibility of applications and the extent of information they present. This choice must be modified each time users transition between contexts, such as changing tasks or environments. Considering the frequent occurrences of context switches throughout the day, introducing automation to MR interfaces becomes essential to address this challenge. A real-time optimization-based approach could be developed by inferring the user's current cognitive load using gaze metrics (for example, the Index of Pupillary Activity \cite{duchowski2018index}) and knowledge of their task and surroundings \cite{lindlbauer2019context}. This approach would enable automatic control over the timing, location, and amount of information displayed by applications.

Implicit gaze effectively measures user confidence while manually mid-air typing on a virtual keyboard in VR \cite{gupta2023}. For instance, postural reinforcement haptic feedback reduced expert QWERTY typists' visual attention to the mid-air keyboard by 44\%, as shown by increased comfort, higher angular distance between gaze and key, and reduced gaze switching. Gaze metrics offer an objective way to quantify subjective user ratings, providing an alternative perspective on user confidence during typing.

Gaze metrics may also be employed to estimate the user's current skill level and assess performance \cite{chew2017skill}. Essentially, gaze data can be utilized to create metrics related to expertise \cite{wang2022algorithmic}, enabling the tracking of learning progression \cite{ diaz2017gaze}. Establishing gaze-based metrics that identify trends in user behavior during XR experiences can contribute to an expertise engine. This engine interacts with the user and system, providing feedback to enhance skill levels and overall experience. For example, increased expertise might require adjusting feedback levels or switching modalities. By integrating expertise-based gaze metrics, the expertise engine facilitates these adaptations, improving the user's skill level and technology on-boarding experience.

\subsection{Application: Using Implicit Gaze And Context to Optimize User Interface Prior To Interaction}
Contextual and Adaptive UIs provide needle-in-the-haystack interfaces, most aligned with a user's context, in a timely manner for users to interact with without exerting willpower to bring up that interface. A user's context relates to the user's current environment. It could include the time, location, task being performed, objects in the environment and the state of these objects, and many more.  Context and gaze analysis enhance interaction efficiency by preloading content and displaying pertinent UI elements before user engagement. By assessing the user's context, and considering factors like task, location, and environment, the system can anticipate needs and load relevant content. For example, given the location of the aisle the user is in and given other contextual factors such as user selection history, time of the day etc, the system can preload and adjust the shopping list to reflect the most relevent items for the user at that moment. Gaze behavior, tracking eye movements and fixations, offers insights into attention and intention \cite{mahanama2022eye, hadnett2019effect}. Recognizing areas of interest allows a prioritized display of relevant UI elements. For example, if the user fixates on an object in the world, the system can dynamically load and make it interactable, anticipating it as the intended target \cite{tonnis2014boundary,tonnis2014placing}. This proactive approach reduces cognitive load since the the user is given the right amount of information at the right time without being overloaded, streamlining access to information and actions, ensuring a smoother and more efficient user experience without delays or additional searching.

\section{Operating gaze in explicit and implicit modes} 
Head mounted devices today tend to constrain users to use their eyes primarily as a targeting mechanism, which, as described above, is an explicit form of input. While the industry is still early in leveraging gaze for spatial interactions, there are multiple advantages of having systems utilize the implicit mode of eye gaze and we discuss them here.

\subsection{Recommendations for Pushing XR Gaze Research From Explicit to Implicit Mode}

If we were to iterate on the current explicit state of gaze targeting, and employ the implicit mode of gaze, we can design systems to be even better aligned with how people use their eyes to interact with technology. We use our eyes naturally for scanning, visual search, navigation, and joint attention, among other tasks\cite{trepkowski2019effect,foulsham2011and,olk2018measuring,kredel2017eye}. In the existing platforms of human-computer interaction such as personal and mobile computing, the eyes play an implicit role while manual selections are executed through button presses, like on a keyboard or through touch gestures. Analogously, in the newly emerging spatial computing platform, given the potential limitations with using gaze explicitly, we strongly advocate future gaze interactions to leverage existing technologies to understand better how we can leverage natural gaze movements.  As shown earlier, not all eye movements are conscious. Our eyes tend to explore things and are mapped to many other activities (thinking, attraction) that heavily depend on context. Using natural gaze behavior to estimate a user's behavioral state and using this information to enhance user experience will allow users to use their gaze naturally thus decreasing physical and mental exertion.

We receive a lot of information from what people naturally do using their eyes, which is powerful when providing them with what they want. At the same time, these are also closely coupled with events happening contextually. As we've described before, this gives insights into users' state of mind, and doing so enables technology to serve their needs best. While using gaze as an explicit targeting modality is a good first step, it is essential to realize that there is so much more that the eyes have to offer in naturalistic HCI if we were to capitalize on the implicit tendencies of our eyes. 

In the past, research was constrained to devices with eye trackers but lacked world-facing sensors. In limited scenarios,  groups collected video-based world information combined with eye tracking, but such resources were scarce. Now, with the advent of wearables featuring eye trackers and context-sensing capabilities, researchers can gain deeper insights into natural gaze behavior. Utilizing these devices for large-scale research, such as Aria program glasses \cite{somasundaram2023project}, allows capturing naturalistic gaze data during users' daily activities. These could include daily tasks such as going on a typical walk, taking a bike ride, cooking or gardening.  These open up a plethora of research opportunities for the larger research community to act on. Research explorations could look deeper into cognitive load, scanpaths across objects \cite{burlingham2024motor, basu2021neural}, and much more as users perform various tasks and do so in scale. Before applying gaze algorithms, it's essential to grasp human behavior in captured scenes. With this understanding, AI advancements enable understanding expected behaviors, considering hand and eye coordination. Segmentation of objects, environments, and scenes provides crucial contextual information. While these are naturalistic gaze interactions in the physical world, virtual interfaces introduce new possibilities, such as studying gaze patterns during tasks like cooking with a virtual UI, or riding a bike with virtual directional instructions. This is a potential area of future research -- how do people now perform tasks and how how does gaze data and behavior change with virtual UI placed in physical or virtual environments? Furthermore how does gaze behavior change with more automation in people's day to day life? As contextual apps with gaze integration proliferate, iterative development of implicit gaze research becomes feasible over time.

\subsection{Challenges for Accommodating Different Modes of Interactions}
Although eye tracking can be applied in both explicit and implicit modes, integrating interactions that utilize eye gaze for these two modes presents challenges to be addressed. For instance, implicit gaze-based models, which rely on capturing reflexive reactionary behavior \cite{Naveen2022}, might mis-classify intentional overt movements made by users during explicit UI control. This can lead to misinterpretations and inaccurate predictions of user intent. As eye gaze becomes increasingly vital in the XR future, serving as a valuable data source for AI or as an interaction modality, it becomes crucial to distinguish between epochs when users are employing gaze implicitly or explicitly. This classification allows for the appropriate triggering of models tailored to the specific interaction mode. 

To address this and other challenges, advancements in eye tracking technology and machine learning algorithms are necessary. The development of sophisticated algorithms capable of accurately discerning between explicit and implicit gaze usage can greatly enhance the effectiveness of XR interactions. By precisely identifying when users are engaging in explicit targeting versus when their gaze behavior reflects implicit behavior, tailored models and interaction approaches can be implemented accordingly. One such approach is to learn the aligned embeddings between gaze and hand movements so that we can use this knowledge of eye-hand coordination to classify implicit vs explicit gaze behavior. Moreover, creating intuitive UIs that seamlessly transition between explicit and implicit gaze-based interactions can further improve the user experience. This includes considering an explainable AI (XAI) approach, such as providing visual cues or feedback to inform users about the mode of gaze interaction they are currently employing \cite{xu2023xair}. Clear indications help users understand how their gaze behavior is being interpreted and ensure that their intentions are accurately captured. Of course, it is also important to investigate privacy protection opportunities, and to assess whether any aspects of the user's state raise concerns and require additional privacy-protection measures.

\section{Conclusion}
In this paper, we introduced how eye tracking can be used in different ways in XR to improve user experience. Through a comprehensive discussion on the state-of-the-art technology for explicit and implicit modes of gaze interaction, we encourage researchers to accelerate research leveraging implicit gaze as a means to understand user intent and state. While explicit gaze is a first step to interactions using eye tracking, implicit gaze allows systems to grasp context intuitively from the user's perspective. This approach aligns with the natural eye usage patterns of users, empowering systems to deliver precise responses and interfaces tailored to their immediate context. Systems that are capable of discriminating implicit gaze tasks gain a deeper understanding of users, marking a crucial step towards contextual interfaces. As research systems evolve to capture gaze data with users in various environments and contexts, they provide many unique opportunities for research to evaluate how users naturally use gaze in their day to day, and this will enable an iterative approach to build future implicit interfaces.

\bibliographystyle{ACM-Reference-Format}
\bibliography{sample-ceur}

\end{document}